\documentclass[aps,prl,epsf,twocolumn,groupedaddress]{revtex4}
\usepackage{graphicx}%
\begin{document}


\title{Contacting single bundles of carbon nanotubes with alternating electric fields
}


\hyphenation{nano-tubes}

\author{R. Krupke}
\email[]{ralph.krupke@int.fzk.de}
\author{F. Hennrich}
\author{H. B. Weber}
\author{D. Beckmann}
\author{O. Hampe}
\author{S. Malik}
\affiliation{Forschungszentrum Karlsruhe, Institut f\"ur Nanotechnologie, D-76021 Karlsruhe}

\author{M. M. Kappes}
\affiliation{Forschungszentrum Karlsruhe, Institut f\"ur Nanotechnologie, D-76021 Karlsruhe\\
Universit\"at Karlsruhe, Institut f\"ur Physikalische Chemie I, D-76128 Karlsruhe}

\author{H. v. L\"ohneysen}
\affiliation{Universit\"at Karlsruhe, Physikalisches Institut, D-76128 Karlsruhe\\
Forschungszentrum Karlsruhe, Institut f\"ur Festk\"orperphysik, D-76021 Karlsruhe}


\date{\today}

\begin{abstract}
Single bundles of carbon nanotubes have been selectively deposited from suspensions onto sub-micron electrodes with alternating electric fields. We explore the resulting contacts using several solvents and delineate the differences between Au and Ag as electrode materials. Alignment of the bundles between electrodes occurs at frequencies above 1 kHz. Control over the number of trapped bundles is achieved by choosing an electrode material which interacts strongly with the chemical functional groups of the carbon nanotubes, with superior contacts being formed with Ag electrodes.
\end{abstract}


\maketitle
Carbon nanotubes have been demonstrated both theoretically and experimentally to be quasi 1-dimensional solids with unique electronic properties \cite{phys}. They can be either metallic or semi-conducting depending on their diameter and/or helicity of the arrangement of the graphitic rings in their wall. Metallic nanotubes, which appear to be ballistic conductors on a length scale up to a micron with critical current densities of 10$^9$ A/cm$^2$, are ideal nanoscale wires \cite{yao}, whereas semiconducting nanotubes enable to build nanoscale field effect transistors \cite{tans}. Furthermore, superconductivity has been reported for metallic nanotubes \cite{kociak}.

A major problem in the realization of electronic circuits is the difficulty to wire-up carbon nanotubes, i.e. to position and contact them in a controlled way. This is an essential requirement for reliable results. 
Several methods have been reported so far, including: (a) spraying of nanotubes \cite{tans,bockrath} - a method where nanotubes are deposited in a random manner onto silicon prior or after lithographic structuring of metallic contacts, (b) catalytic growth of nanotubes \cite{kong} - a high temperature process where nanotubes are grown on silicon from predeposited catalyst islands, and (c) self-assembling on chemically modified surfaces \cite{liu} - a process using chemically modified silicon surfaces for the selective deposition of carbon nanotubes.

Up to now, however none of these methods has been efficient enough to supersede the others. Recently it has been shown that large numbers of bundles of carbon nanotubes can be aligned upon deposition from dispersion by applying an alternating electric field to an electrode assembly patterned onto a silicon surface \cite{chen}. The effect is thought to be due to their large and anisotropic electronic polarizability. In order to use this effect as a viable method for wiring up nanotubes into circuits, one needs to have control over the number of trapped nanotubes.

We show here that it is possible to trap a {\it single} bundle of carbon nanotubes onto electrodes with the use of alternating electric fields by choosing an electrode material which interacts strongly with the chemical functional groups of the carbon nanotubes. 

Single-walled carbon nanotubes were grown in a laser ablation system \cite{lebedkin}. The as-grown material contains not only nanotubes but also catalyst particles and amorphous carbon, both of which can be removed to a large degree by a weak acid treatment \cite{reflux}. Thermogravimetric analysis and transmission electron microscopy yield a 99$\%$ sample purity. Remaining impurities are catalyst particles embedded in carbon shells, separate from the nanotubes \cite{malik}. The nanotube bundles, usually several microns long, were shortened with concentrated HNO$_3$ / H$_2$SO$_4$ (1:3) for 3 h at 60$^\circ$C. Finally the shortened tubes were suspended in N,N-dimethylformamide (DMF).

UV-VIS and IR spectra of the suspended tubes are in agreement with literature reports of metallic and semi-conducting nanotubes having COOH functional groups associated with the acid treatments \cite{chiang}. For the trapping experiments, the suspension was diluted to the extent that the liquid appears colorless and transparent (nanotube concentration $\approx$ 10 ng/ml).

Electrodes were prepared on thermally oxidized silicon substrates with standard electron beam lithography. The thickness of the oxide layer is 600 nm. The electrodes are 20 nm thick, 80-150 nm wide, and the electrode distance is of the order of 100 nm. Au and Ag are used as top electrode materials.

Prior to trapping of tube bundles, the structure has been bonded onto a chip carrier and wired up with a series resistance $R_S = 500$ M$\Omega$. The circuit is powered by a low-impedance frequency generator with optional dc-offset voltage. During trapping, the electric current is monitored by the ac or dc voltage across the series resistance, $V_{AC}$ and $V_{DC}$, respectively. The $V_{DC}$ has been measured with a high-impedance voltmeter. For the measurement of the $V_{AC}$ we have chosen a lock-in amplifier instead of an ac voltmeter due to its higher input impedance of $R = 10$ M$\Omega$ and $C = 30$ pF.

After switching on the frequency generator, a drop of nanotube suspension ($\approx 10$ $\mu $l) is applied onto the chip with a pipette. After a delay of typically one minute, the drop is blown off the surface with nitrogen gas. Finally the generator is turned off and the sample is subjected to SEM characterization and transport measurements.
 \begin{figure}
   \includegraphics[width=20pc]{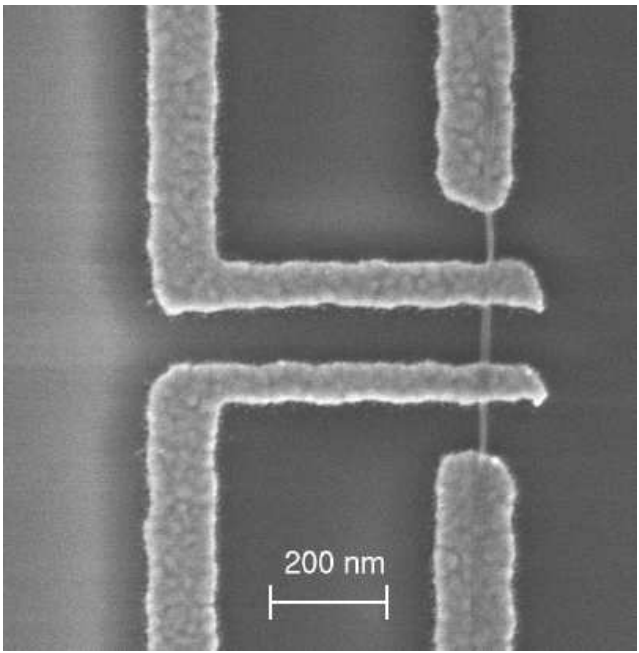}%
   \caption{Scanning electron micrograph of a single bundle of carbon nanotubes trapped on four Au electrodes. The alternating electric field has been generated between the upper right and lower right electrode. The other two electrodes were at floating potential. The bundle diameter is 9 nm.}
 \end{figure}

Fig. 1 shows an example of a single carbon nanotube bundle trapped along four Au electrodes using the experimental setup described above. The trapping has been performed with an applied field of $V_{rms} = 1$ V at the frequency $f = 1$ MHz. The chip was exposed to the tube suspension for 20 s.

Apparently, the bundle has been trapped with an excellent alignment and this result is highly reproducible at this frequency, independent of the chosen solvent or electrode material. We have observed in our experiments that nanotubes align only at trapping frequencies above 1 kHz. At lower frequencies nanotubes attach to electrodes with random orientation. This is consistent with reports of a better alignment at higher frequencies \cite{chen,yamamoto}.
A plausible explanation for the observations is that polarisable nanotubes with random orientation with respect to a static electric field acquire induced dipole moments pointing mainly along their axes \cite{benedict}. Hence the electric field of two opposing electrodes aligns nanotubes along the field lines. The inhomogeneity of the electric field generates, in addition, an attractive force towards the central area between the electrodes. Finally, nanotubes attach to the electrodes and bridge them via a straight line.

In our experiment the alignment works with alternating fields at higher frequency only ($f > 1$ kHz), although the whole frequency range used is quasi static for the electronic system of the nanotubes. We propose that this is due to effects of the suspension: (1) at low frequencies a Helmholtz double layer forms at each electrode, weakening the electric field. (2) nanotubes are negatively charged in DMF \cite{liu}, therefore the suspension is partly of ionic nature. At low frequency the ions are able to follow the alternating polarization of the nanotubes and can shield them effectively. This also applies for any residual ions left from the acid treatment.

 \begin{figure}
   \includegraphics[width=20pc]{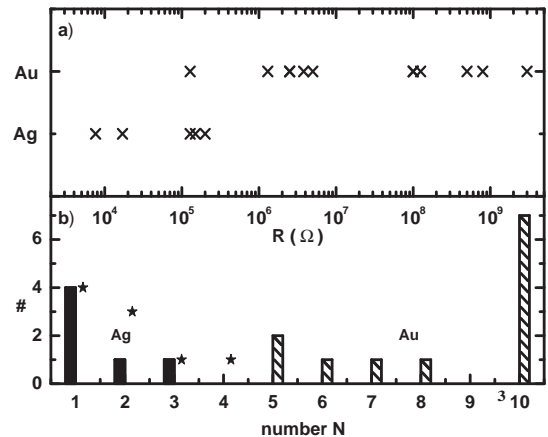}
   \caption{(a) The two-terminal resistance of single bundles on Au and Ag. (b) The number of trapped bundles per experiment on Au and Ag. Trapping experiments on Au with very short exposure time ($< 20$ s) are depicted by (*).}
 \end{figure}

What is the number of trapped bundles? In Fig. 1 only one bundle is trapped, while with Au electrodes we most often observe more than one. In the following we show that the control over the number of trapped bundles depends decisively on the electrode material. The number of bundles $N$ can be reliably determined with SEM. With Au electrodes, $N$ increases with the time interval, the electrodes are exposed to the suspension, and furthermore with the concentration of the suspension. With Ag electrodes, on the other hand, the situation is different. Here, only very few bundles, and quite often just a single one, are trapped independent of time and concentration. Fig. 2b shows the distribution of the number of trapped bundles per experiment.

This observation goes along with a significant reduction and reproducibility of the two-terminal resistance of single bundles when using Ag instead of Au electrodes, as is shown in Fig. 2a.
For an understanding it is necessary to consider the experimental setup in some detail. The series resistance is expected to function as a voltage divider and current limiter. As soon as a contact is formed between the electrodes via a nanotube and the resulting resistance $R_T$ is smaller than the series resistance $R_S$, the applied voltage will mainly drop along the latter. Hence the field between the electrodes will collapse and the trapping of additional tubes be automatically prevented.

This protocol apparently works well only when we use Ag electrodes. The failure of the mechanism with Au electrodes, can be understood in the following way: If an alternating voltage is applied, the voltage divider does operate only at resistance values much lower than the series resistance because the unavoidable capacitive reactance  Im $Z = \omega C$ of the leads and of the measurement devices, shortcircuits the series resistance with a typical capacitance of $C$ = 20 pF. For instance, when attempting trapping at $f = 1$ MHz the accumulation of bundles continues, unless a resistance of the order of $R_T = 10$ k$\Omega$ or lower is formed. For lower frequencies $R_T$ is correspondingly shifted to larger values.

Although low two-terminal resistances have been observed for metallic single-walled carbon nanotubes \cite{nygard}, still many authors report values around 100 k$\Omega$ to 1 M$\Omega$, especially for nanotubes positioned on top of the electrodes \cite{yao,bockrath}. In our experiments the two-terminal resistance of single bundles of nanotubes is consistently smaller for contacts formed with Ag electrodes than with Au electrodes (Fig. 2a). These resistance measurements have been performed after removal of the solvent, and it is unclear whether a difference in resistance already shows up in the presence of just the solvent. We have therefore made in-situ measurements during trapping for samples with Au and Ag electrodes.

\begin{figure}
   \includegraphics[width=20pc]{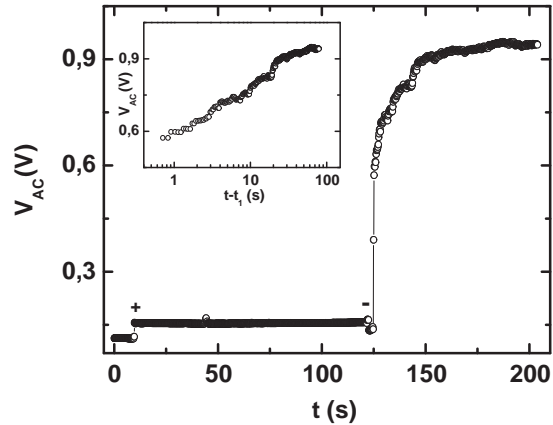}
   \caption{Evolution of $V_{AC}$ across the series resistance during trapping on Au electrodes at $f = 1$ kHz. At $t_0$ (+) the suspension is applied and at $t_1$ (-) the suspension is blown off. The inset shows the data for $t > t_1$ on a semi logarithmic scale.}
 \end{figure}

Fig. 3 displays the alternating voltage drop $V_{AC}$ at the series resistance during trapping the nanotubes on Au electrodes. The first small increase of $V_{AC}$ marks the time when the nanotube suspension was applied (+). This increase is attributed to the finite conductivity of DMF and does not indicate instantaneously trapped nanotubes. During the time the suspension wets the chip, no significant additional increase of $V_{AC}$ can be seen. After blowing off the suspension (-) $V_{AC}$ starts to rise steeply, after a short initial drop. The data clearly shows that an electric contact on Au is formed only after removal of the suspension, which is consistent with the observation that the number of nanotubes on Au is not controllable with our experimental setup.

Similar measurements during trapping on Ag electrodes are shown in Fig. 4a-b. Here a small dc bias is added to the ac signal and detected with a high impedance voltmeter. The first increase of $V_{DC}$ marks the moment when the nanotube suspension is applied (+), similar to the effect on $V_{AC}$ discussed above. After some time a sudden increase of $V_{DC}$ is observed (o), which indicates the trapping of a nanotube bundle. Finally the suspension is blown off (-).

Clearly an electric contact has been established between nanotubes and Ag already in the presence of the suspension. Hence, with Ag electrodes the number of nanotubes is controllable with our experimental set-up. To corroborate the difference of the two electrode materials, we show in Fig. 4c another example of trapping on Au electrodes, this time with $V_{DC}$. Again, it is only after removing the suspension that an electric contact is formed. It is important to note that we have observed a similar trapping behavior to Au electrodes with other solvents like water, isopropanol, cyclohexane and 1,2-dichlorobenzene.

\begin{figure}
   \includegraphics[width=20pc]{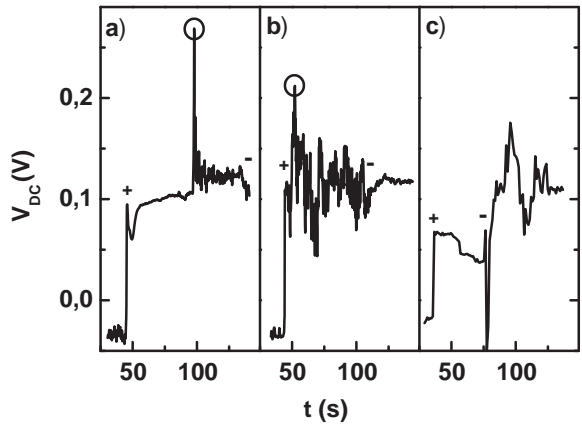}
   \caption{Evolution of $V_{DC}$ across the series resistance during trapping on Ag electrodes (a,b) and Au electrodes (c) at $f = 1$ MHz. The moment when the suspension is applied and blown off, is marked with (+) and (-) respectively. The circles mark the point of time, where a bundle of carbon nanotubes has been trapped.}
 \end{figure}

The question arises why nanotubes form an electric contact with Ag in the presence of a suspension and why they do not with Au. The origin may be found in the strong affinity of the COOH groups of our acid-treated nanotubes to Ag surfaces. For instance it is known that n-alkanoic acid [CH$_3$(CH$_2)_m$COOH] forms a self-assembled monolayer on native Ag oxide surfaces \cite{tao} and that short, COOH-functionalized nanotubes attach parallel to the surface normal of Ag films \cite{wu}. Hence it is quite likely that COOH-nanotube bundles undergo chemical bonding to Ag electrodes even in the presence of solvent.

In contrast, COOH groups have no affinity to Au surfaces. From the steep increase of $V_{AC}$ in Fig. 3 we suspect, that an adsorbed layer of solvent inhibits the formation of an electrical contact in liquid. We have analyzed the time dependence of $V_{AC}$ after blowing off the DMF. $V_{AC}$ is directly proportional to the conductance of the Au-nanotube-Au structure for $V_{AC} \ll V_{rms}$. The inset of Fig. 3 demonstrates that $V_{AC}$ increases roughly logarithmically with time. Such logarithmic rate laws are known from chemisorption and oxidation processes \cite{landsberg}, and in this case is probably due to desorption of the DMF layer between nanotubes and Au electrodes. We have carried out the following control experiment to exclude a possible doping of nanotubes by DMF as known for oxygen, air or electrolytes \cite{collins,kruger}: Once the DMF solvent has been blown off, the resistance remains stable even after subsequent readmittance of solvent. Hence the $V_{AC}$ behavior after removal of the solvent supports our notion of a solvent layer inhibiting a low-resistance contact for Au during trapping.

In summary we have demonstrated that a single bundle of carbon nanotubes can be selectively trapped onto electrodes with alternating electric fields. The alignment is excellent for frequencies larger than 1 kHz. Matching of the electrode material with the chemical functional groups of the carbon nanotubes appears to be essential for controlling the number of trapped bundles and for obtaining low contact resistance. Best results for nanotubes with COOH groups have been achieved with Ag electrodes. With Au electrodes the formation of an electrical contact during trapping is inhibited, probably by a layer of solvent. Thus, besides establishing the use of alternating electrical fields to position nanotubes, our results stress the important role of chemical bonding in actually forming the contact between metals and nanotubes. We expect this method to work also with isolated nanotubes (not bundles), which were not available in our experiments.
\begin{acknowledgments}
The authors would like to thank S. Lebedkin, R. Ochs and J. Reichert for discussions. This work was partly supported by the Deutsche Forschungsgemeinschaft through Sonderforschungsbereich 551.
\end{acknowledgments}

\bibliography{ralphprl}

\end{document}